# AI Adoption Among Teachers: Insights on Concerns, Support, Confidence, and Attitudes

Vanessa Baltazar Sibug
College of Education
Pampanga State University
Bacolor, Pampanga, Philippines
vbsibug@pampangastateu.edu.ph

Maria Anna David Cruz
College of Hospitality and Tourism Management
Pampanga State University
City of San Fernando, Pampanga
Philippines
madcruz@pampangastateu.edu.ph

Vicky Pineda Vital
College of Education
Pampanga State University
Mexico, Pampanga, Philippines
vpvital@pampangastateu.edu.ph

Juvy Cruz Grume
College of Computing Studies
Pampanga State University
Bacolor, Pampanga, Philippines
jncruz@pampangastateu.edu.ph

Almer Balingit Gamboa
College of Education
Pampanga State University
Mexico, Pampanga, Philippines
abgamboa@pampangastateu.edu.ph

Emerson Quiambao Fernando
College of Education
Pampanga State University
Lubao, Pampanga, Philippines
eqfernando@pampangastateu.edu.ph

Lloyd David Feliciano
College of Education
Pampanga State University
Lubao, Pampanga, Philippines
ldfeliciano@pampangastateu.edu.ph

Jordan Lansang Salenga
College of Computing Studies
Pampanga State University
Bacolor, Pampanga, Philippines
jlsalenga@pampangastateu.edu.ph

John Paul Palo Miranda*
College of Computing Studies
Pampanga State University
Mexico, Pampanga, Philippines
jppmiranda@pampangastateu.edu.ph

## Abstract

The study examines the adoption of artificial intelligence (AI) tools in education by analyzing the roles of institutional support, teacher confidence, and teacher concerns. It aims to determine whether teacher concerns moderate the relationship between institutional support and two outcomes: teacher confidence and attitudes toward AI adoption. The sample included 260 teachers from the Philippines. Composite scores were calculated for institutional support, confidence, concerns, and attitudes. Moderated multiple regression analysis showed that institutional support significantly predicted both teacher confidence and attitudes toward AI. However, teacher concerns did not significantly moderate these relationships. A follow-up mediation analysis tested whether confidence explains the effect of institutional support on attitudes. Results showed full mediation. The indirect effect was significant based on the Sobel test, and the direct effect became non-significant when confidence was included in the model. This shows that institutional support improves teacher attitudes by increasing their confidence. The study recommends that institutions provide structured and ongoing support to strengthen teacher confidence. Professional development, mentoring, and AI integration in teacher education programs can increase readiness and support effective AI adoption.

*Corresponding author

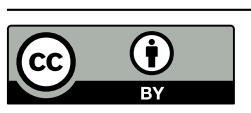

## CCS Concepts

**• Applied computing**; **• Education**; **• Social and professional topics**; **• Professional topics**; **• Human-centered computing**; **• Collaborative and social computing**; **• Empirical studies in collaborative and social computing**;

## Keywords

Teacher confidence, Institutional support, Technology integration, Technology adoption, AI tools

## 1 Introduction

Teachers guide students, promote critical thinking, and support ethical reasoning in increasingly technology-driven classrooms [4]. As artificial intelligence (AI) tools become integrated into education, teachers must adapt by building digital skills, collaborating with institutions, and balancing innovation with meaningful human interaction. AI supports personalized learning and adaptive assessments but also brings technical and ethical challenges [5]. To address these, teachers must build competencies that allow them to use AI effectively while maintaining pedagogical goals [13].

This study examines how institutional support, teacher confidence, concerns, and attitudes influence the adoption of AI tools in education. Using moderated multiple regression (MMR), it tests whether concerns moderate the effects of institutional support on confidence and attitudes. The study aims to identify key factors that promote teacher readiness for AI integration and contribute to research on technology adoption in education. Grounded in the Unified Theory of Acceptance and Use of Technology (UTAUT) [1], it defines institutional support as a facilitating condition, confidence as self-efficacy, concerns as inhibitors, and attitudes as behavioral intention. The analysis extends UTAUT by exploring how institutional and psychological factors shape AI adoption in teaching.

### 1.1 Problem Statement

Despite the growing use of AI in education, many teachers struggle with adoption due to limited training, unreliable access to technology, and concerns about ethics and data privacy [2, 6, 9, 11]. While research confirms the influence of institutional support and teacher confidence on technology use, few studies explore how these factors interact to shape attitudes toward AI. This study addresses that gap by testing whether teacher concerns weaken or change the effect of institutional support on confidence and attitudes, and whether confidence explains how support leads to more positive attitudes. Clarifying these relationships helps identify where schools should focus efforts to increase teacher readiness for AI adoption.

## 2 Method

To investigate the factors influencing AI adoption, this study employed a quantitative research design to examine the relationships among teacher concerns, institutional support, confidence, and attitudes toward AI adoption. Moderated multiple regression (MMR) tested whether concerns moderated the effects of support on confidence and attitudes. This method allowed for the analysis of the four constructs and their relationships. Data analysis was conducted using Python libraries and Jupyter Notebook. A mediation analysis followed to test whether teacher confidence mediated the relationship between support and attitudes, using the Baron and Kenny method and the Sobel test to evaluate the indirect effect.

Purposive sampling ensured representation across various educational levels and teaching disciplines. The sample included 260 teachers from Pampanga, Philippines: 52 elementary, 161 secondary, and 47 college educators. Data collection involved both pen-and-paper surveys and Google Forms. Permissions were secured from school heads and supervisors prior to data collection. The survey instrument contained 35 items measuring the four constructs using a five-point Likert scale (1 = strongly disagree to 5 = strongly agree). Content validity was assessed by two educational technology experts who evaluated item alignment with theoretical constructs. Reliability analysis confirmed strong internal consistency across all constructs ($\alpha > 0.70$). The demographic profile showed that over 75% of participants were female, and approximately 80% were licensed professional teachers. Ages ranged from 22 to 59 years, and teaching experience ranged from 1 to 36 years.

## 3 Results and Discussion

Teachers demonstrated strongly positive attitudes toward AI tools (mean = 4.92, SD = 0.86) with broad agreements on the benefits of AI in enhancing learning, engagement, and personalization. Concern regarding AI adoption was low (mean = 2.14, SD = 0.89) which suggests that minimal apprehension regarding technical issues, data privacy, and potential overreliance on AI. High confidence is reported particularly in their ability to utilize AI tools (mean = 4.72, SD = 0.89). However, confidence levels pertaining to handling technical issues were comparatively lower which highlights an area that may require additional training and support. Institutional support for AI integration was generally strong (mean = 4.32, SD = 1.08) particularly in terms of access to devices, software, and ethical guidelines. Nevertheless, the reported moderate support for internet reliability and infrastructure suggests room for further institutional improvement.

The first model examined whether institutional support significantly influences teacher confidence and whether this relationship is moderated by concerns. The results revealed a strong, statistically significant positive effect of support on confidence ($\beta = 0.537$, $p < 0.001$), indicating that increased access to resources, training, and infrastructure enhances teacher confidence in adopting AI. However, concerns did not significantly moderate the relationship between support and confidence ($\beta = 0.021$, $p = 0.661$), nor did they have a direct effect on confidence ($\beta = -0.104$, $p = 0.625$). This may indicate that institutional factors such as access to resources and training outweigh individual apprehensions in shaping confidence. It is also possible that the participants had already formed positive attitudes toward AI which reduced the perceived impact of concerns. These findings suggest a normative shift toward AI acceptance particularly among teachers exposed to structured institutional programs. The model accounted for 47.6% of the variance in confidence ($R^2 = 0.476$) and demonstrated strong overall fit ($F = 87.35$, $p < 0.001$). These findings reinforce that irrespective of their concerns, institutional support is a central determinant of teacher confidence in adopting AI. Schools must ensure access to professional development, adequate resources, and technical assistance to address barriers to implementation. The limited effect of concerns on confidence indicates that teachers are more willing to adopt AI when they perceive strong institutional backing, regardless of potential risks. These findings align with previous research emphasizing the importance of infrastructure, pedagogical support, and belief in technology's benefits [14], as well as the roles of organizational culture and administrative support in promoting technology integration [12]. In contrast, a lack of institutional support can hinder adoption even when teachers acknowledge the value of technology [15]. Comprehensive support which includes infrastructure maintenance and teacher motivation which remains critical for successful AI integration and improved educational outcomes [8].

The second model analyzed the relationships among institutional support, concerns, and attitudes while controlling for the effect of confidence on attitudes. The confidence had a strong and statistically significant positive effect on attitudes ($\beta = 0.855$, $p < 0.001$), indicating that teachers with higher confidence in using AI are more likely to adopt favorable attitudes. The direct effect of support on attitudes was not significant ($\beta = 0.128$, $p = 0.106$), suggesting that support alone does not directly shape attitudes. Concerns also did not significantly predict attitudes ($\beta = 0.220$, $p = 0.151$) and did not moderate the support-attitude relationship ($\beta = -0.047$, $p = 0.173$). The model explained 74.2% of the variance in attitudes ($R^2 = 0.742$) and showed a strong overall fit ($F = 205.9$, $p < 0.001$). Institutional support indirectly influences attitudes through its effect on confidence, while concerns do not exert a significant influence. This highlights the importance of initiatives that build confidence among teachers, as this factor has the most significant impact on their readiness and behavior to adopt AI [3, 7]. Schools and educational institutions should focus on creating robust support systems such as training programs, access to technical resources, and well-established infrastructure to enhance teachers' confidence and preparedness for AI integration [11].

Building on these results and to further explore the underlying mechanism of these relationships, the mediation analysis revealed that institutional support significantly predicted teacher confidence ($\beta = 0.58$, $p < .001$), and in turn, confidence strongly predicted attitudes ($\beta = 0.85$, $p < .001$). However, when confidence was included in the model, the direct effect of institutional support on attitudes became non-significant ($\beta = 0.03$, $p = .385$) which indicates a full mediation. The indirect effect of institutional support on attitudes through confidence was statistically significant ($a \times b = 0.4961$), as confirmed by a Sobel test ($z = 12.63$, $p < .001$). This suggests that institutional support enhances teacher attitudes toward AI adoption primarily by fostering confidence [6], rather than exerting a direct effect. Confidence acts as a key psychological mechanism that converts institutional inputs into favorable perceptions. This aligns with prior work on self-efficacy in technology adoption which emphasized the role of personal agency over external concerns in shaping behavioral intentions. Furthermore, it reinforce the critical role of confidence as a psychological mechanism that translates support into positive dispositions toward educational technology [10].

## 4 Conclusion and Recommendations

This study investigated how institutional support, teacher confidence, and concerns affect the adoption of AI tools in education. Results from MMR demonstrated that institutional support significantly boosts teacher confidence and positive attitudes toward AI adoption, while teacher concerns did not moderate these effects. A subsequent mediation analysis confirmed that teacher confidence fully mediated the relationship between institutional support and positive attitudes, indicating that confidence is critical to readiness for AI integration. This study recommends that educational institutions should focus on enhancing teacher confidence through targeted professional development programs emphasizing self-efficacy, adaptability, and pedagogical skills. Continuous access to resources, technical assistance, structured mentoring, and incorporating AI literacy in teacher education programs are essential. Policymakers and school leaders should recognize that resistance to AI often reflects systemic support gaps rather than individual reluctance. Building supportive institutional environments that foster teacher confidence is vital for responsible, sustainable, and effective AI adoption in education.